\documentclass[fleqn,10pt]{wlscirep}
\usepackage[utf8]{inputenc}
\usepackage[T1]{fontenc}
\usepackage{rotating}
\title{Physics-Guided Deep Learning For High Resolution X-ray Imaging}

\author[1,*,+]{Shao Xian Lee}
\author[1,+]{Aashwin Ananda Mishra}
\author[1]{Ariel Arnott}
\author[1]{Meriame Berboucha}
\author[1]{Nina Boiadjieva}
\author[1]{Gourab Chatterjee}
\author[1]{Eric Cunningham}
\author[1]{Nick Czapla}
\author[1]{Gilliss Dyer}
\author[1]{Jonathan Ehni}
\author[1]{Robert Ettelbrick}
\author[4]{Anna Grassi}
\author[4]{Mickael Grech}
\author[1]{Philip Hart}
\author[1]{Dimitri Khaghani}
\author[1]{Hae Ja Lee}
\author[1]{Peregrine McGehee}
\author[1]{Bob Nagler}
\author[5]{Paul Neumayer}
\author[4]{Caterina Riconda}
\author[1]{Marc Welch}
\author[7]{Andrea Zabala}
\author[1]{Eric Galtier}
\author[2]{Quynh L. Nguyen}
\affil[1]{SLAC National Accelerator Laboratory, 2575 Sand Hill Rd., Menlo Park, CA, 94025, USA}
\affil[2]{NVIDIA Corporation, 2788 San Tomas Expressway, Santa Clara, CA 95051, USA}
\affil[3]{European XFEL, Holzkoppel 4, 22869 Schenefeld, Germany}
\affil[4]{Laboratoire pour l'Utilisation des Lasers Intenses, Sorbonne Universités, CNRS, Ecole Polytechnique, CEA, 75252, Paris, France}
\affil[5]{GSI Helmholtzzentrum für Schwerionenforschung GmbH, Darmstadt, Germany}
\affil[6]{Lawrence Berkeley National Laboratory, 1 Cyclotron Road, Berkeley, CA 94720, USA}

\affil[*]{lsx249@slac.stanford.edu}

\affil[+]{these authors contributed equally to this work}

\begin{abstract}
Imperfections in X-ray imaging systems can limit their performance, especially in High Energy Density (HED) or Inertial Fusion Energy (IFE)-relevant experiments that are typically single shot, by introducing structured, non-stationary features that overlap with the signal of interest. When the X-ray transmission is reconstructed by typical flat-field normalization, even small shot-to-shot drift of structured features imprints residual patterns onto transmission maps, degrading signal visibility and biasing measurements such as electron density, velocity and feature sizes.
We investigate this limitation by modeling the artifacts as a separable feature layer and training a U-Net architecture to estimate and infer them directly from the experimental data. We compare our method against Fourier filtering and more advanced procedures like Dynamic Flat-Field Normalization (DFFN) to evaluate artifact suppression capability and signal preservation in the reconstructed transmission maps. In multiple synthetic injection tests, our Physics-Guided Deep Learning approach is able to obtain an improvement in mean Structural Similarity Index (SSIM) from $0.345$ to $0.906$ and from $0.0679$ to $0.945$, while better preserving filament profiles and reducing degradation of the filament signal during artifact suppression. Additionally, we utilize deep ensembles to obtain predictive epistemic uncertainty estimates for the U-Net based reconstruction, to ensure Out Of Distribution (OOD) robustness for this procedure.

\end{abstract}
\begin{document}

\flushbottom
\maketitle

\thispagestyle{empty}

\section*{Introduction}

Imaging using X-ray free-electron lasers (XFELs) offer high spatial and temporal resolution for studying ultrafast HED and IFE-relevant processes\cite{Galtier2025MXI}. However, obtaining quantitative measurements with resolution better than 200 nm can be challenging. A major limitation in direct X-ray imaging is the presence of structured artifacts. In particular, density inhomogeneities in the beryllium materiel used in the Compound Refractive Lenses (CRL) of the imaging system can introduce non-stationary features that spatially overlap the signal of interest and vary slightly from shot to shot due to X-ray beam pointing jitter, as well as small changes in the XFEL central photon energy and spectral profile\cite{Bradley2002backlighter,Park2008backlighter,Galtier2025MXI}. Other imaging techniques, such as Talbot imaging, do not generally suffer as much from the imaging defects due to the intrinsic filtering of the reconstruction method. However, a compromise must be made between spatial resolution and sensitivity to density gradient in experimental geometry similar than in direct imaging\cite{momose2003talbot}.
Because X-ray transmission maps are typically reconstructed by normalizing a laser-driven image with a flat-field reference, even small shifts in these structured features can prevent clean cancellation and imprint residual patterns onto the reconstructed transmission map\cite{Buakor2022flatfield,Hagemann2021flatfield,VanNieuwenhove2015flatfield,Galtier2025MXI}. These residuals degrade image fidelity and introduce errors in derived quantities such as density, electron velocity, and dimensions of interesting patterns.

This limitation is particularly important for fundamental understanding of high intensity laser/matter  interaction where relativistic particle beams transport in dense plasmas\cite{Dover2025filamentexp,Sawada2024filamentexp}. This process can be studied using a controlled platform, in which a high-power laser irradiates a solid target and drives energetic electrons into the target. The injected electron beam can undergo current-driven instabilities (e.g. the Weibel instability) and break into filamentary channels\cite{Wei2004solidfilament,Romagnani2019solidfilament}. This fundamental knowledge is also of interest for IFE schemes such as fast ignition, where high-power laser pulses generate energetic particles that propagate through solid‑density material and deposit their energy in the compressed fuel to trigger ignition\cite{Kodama2001,Robinson2014,Tabak1994}. Direct imaging of this picosecond-scale filamentation process requires diagnostics with both high temporal and spatial resolution and can be achieved using coherent, ultrashort XFEL pulses\cite{Galtier2025MXI, Schoenwaelder2026filament}.

In direct-imaging geometries, however, the structured feature layer introduced by the imaging system can overlap the filament signal and persist after flat-field normalization. In the ideal case, the transmission is reconstructed as \(T(x,y)=I_{shot}(x,y)/I_{flat}(x,y)\), where \(I_{flat}\) is an X-ray-only image acquired without the sample. When the structured features in \(I_{shot}\) and \(I_{flat}\) are misaligned, this division no longer removes them cleanly, and the residual pattern can obscure filamentary structures or bias the measurement of their extent.

Because the dominant artifact in our data is a structured pattern that drifts between frames rather than purely stochastic noise, conventional denoising methods and homomorphic filtering are ineffective and may even suppress true filament contrast\cite{Oppenheim_1968filtering}. To address this problem, we treat the artifact as a separable feature layer and train a U-Net-based architecture to estimate this feature layer directly from the experimental data\cite{ronneberger2015UNet,Siddique2021UNet,Azad2024UNet}. The predicted feature layer is then removed before transmission reconstruction, yielding transmission maps that are substantially more robust to shot-to-shot artifact drift. Most importantly, once trained, the network requires only a single forward pass to correct each new image. This highly deterministic, millisecond-scale inference time addresses a critical bottleneck at next generation, high repetition rate X-ray facilities, for instance, the diffraction-limited upgrade of the Advanced Photon Source (APS)\cite{fornek2019advanced} and the high repetition rate upgrade of Linac Coherent Light Source (LCLS-II)\cite{LCLSIIHE, mishra2025start,edelen2019machine}. As these beamlines push toward MHz-class repetition rates, the resulting data deluge renders iterative or manually tuned post-processing physically intractable. Our deep learning approach enables real-time, on the fly artifact suppression, making it directly applicable to edge computing data reduction pipelines. Furthermore, because fast inference allows for real-time feature extraction, the network may be coupled with future autonomous experiment steering algorithms. When combined with our deep ensemble uncertainty quantification, the system provides a high-speed ``circuit breaker'', that automatically flags novel physical phenomena, such as shock waves, without interrupting the automated high-throughput data stream.

\section*{Results}

\begin{figure}
    \centering
    \includegraphics[width=1\linewidth]{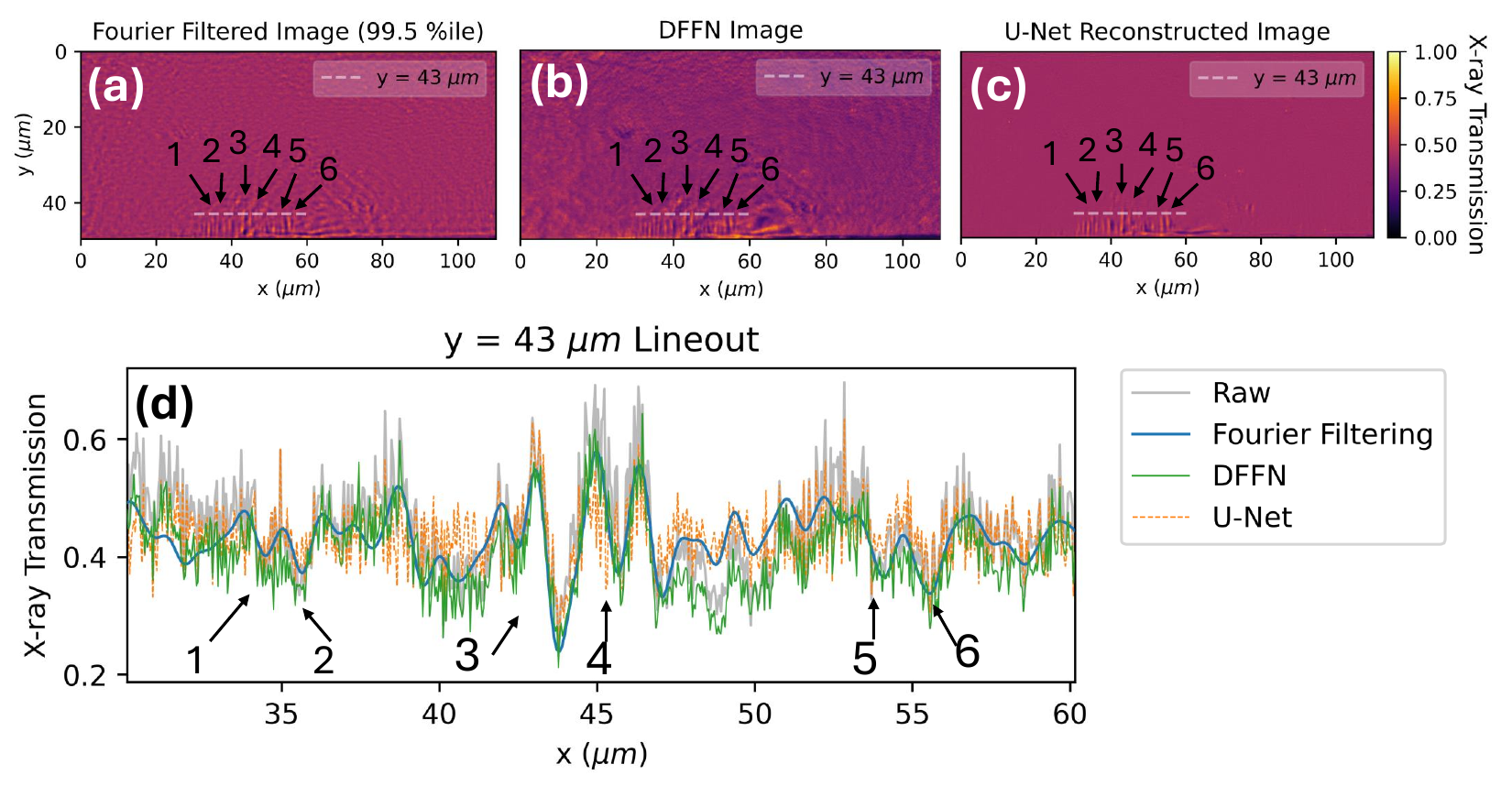}
    \caption{Comparison of defect-suppression methods for the X-ray transmission map. \textbf{(a)} Fourier-filtered image, \textbf{(b)} DFFN image, and \textbf{(c)} U-Net reconstructed image. The dashed line indicates the lineout location at $y=43~\mu\mathrm{m}$ from the sample surface. \textbf{(d)} Corresponding lineouts at $y=43~\mu\mathrm{m}$ from the raw (unprocessed), Fourier-filtered, DFFN, and U-Net reconstructed images. Filaments 1 to 6 are marked in \textbf{(d)}, and their positions are indicated in \textbf{(a)}, \textbf{(b)}, and \textbf{(c)}. The U-Net reconstruction suppresses imaging defects while preserving the filament signal, whereas Fourier filtering attenuates some filament signals (e.g., filaments 1 and 2), and DFFN retains most of the artifacts.}
    \label{fig:Fourier_U-Net}
\end{figure}

We compare and contrast the output of U-Net with two baseline methods: Fourier filtering\cite{Munch2009FFiltering} and Dynamic Flat-Field Normalization (DFFN)\cite{VanNieuwenhove2015flatfield}. The corresponding transmission maps are shown in Fig~\ref{fig:Fourier_U-Net}. In our data, the structured artifacts behave as a multiplicative modulation (attenuation-like) rather than additive, spatially uncorrelated noise. As a result, their spectral content is not cleanly separable from the physical signal. The artifact and filament structures share overlapping spatial-frequency components, so simple frequency masking introduces an inherent trade-off between artifact suppression and filament preservation. Fig.~\ref{fig:Fourier_U-Net}\textbf{(a)} illustrates this limitation by showing the spatial-domain reconstruction obtained by 99.5-percentile filtering. The standard deviation of the reconstructed transmission map outside the filament region of interest (ROI), $\sigma_T$ decreases from $\sigma_T^{raw}=0.0650$ to $\sigma_T^{filtered}=0.0392$. However, the reconstructed component remains speckle-like and contains residual structured texture, indicating that dominant spectral components are not uniquely associated with the artifact layer. Increasing the filtering aggressiveness further suppresses residual patterns but also attenuates filament contrast.

In addition, we applied DFFN, in which the flat-field image is corrected based on the variation in flat-field before transmission reconstruction. In our dataset, this correction modifies the flat field slightly, the resulting transmission map remains visually similar to the raw reconstruction and retains substantial residual artifact structure  (Fig.~\ref{fig:Fourier_U-Net}\textbf{(b)}). The corresponding standard deviation outside the filament ROI is $\sigma_T^{\mathrm{DFFN}}=0.0563$, which suggests that the DFFN does not sufficiently remove the drifting structured imaging artifacts in our data.

Because the feature layer prediction from U-Net is insensitive to the shot-to-shot variation, the same trained model can be applied to both the $I_{shot}$ and $I_{flat}$. After removing the predicted feature layer from each image, we obtained the cleaned images $\tilde{I}_{shot}$ and $\tilde{I}_{flat}$, from which the corrected transmission map is reconstructed as $T_{corr}(x,y)=\tilde{I}_{shot}/\tilde{I}_{flat}$ (Fig.~\ref{fig:Fourier_U-Net}\textbf{(b)}). In contrast to the direct flat-field division, which can imprint residual patterns when the structured artifacts between acquisition of $I_{shot}$ and $I_{flat}$, the U-Net-corrected normalization suppresses these residual artifacts. This yields standard deviation outside the filament ROI, $\sigma_T^{corr}=0.0472$. 

A lineout across the filaments ($y=43\mu m$) is shown in Fig~\ref{fig:Fourier_U-Net}\textbf{(d)} to demonstrate the artifact suppression and preservation of filament profile by U-Net reconstruction, compared to Fourier filtering. This lineout is chosen near the filament ends because accurate identification of the filament tips is important for filament length measurement. The individual filaments are labeled in the lineout profile. As shown in the region between filament 2 and 3, the U-Net reconstruction successfully suppresses the imaging defects, whereas Fourier filtering smooths the profile but fails to suppress the artifacts in this region, and DFFN retains much of the artifacts. For most filaments, all three methods preserve the overall filament profile. However, for some filaments (i.e., filament 1 and filament 2), the Fourier filtering attenuated the filament profile amplitude, which may lead to errors in the filament length measurements, whereas the U-Net reconstruction preserves the filament profile.

\subsection*{Injection Test}

\begin{figure}
    \centering
    \includegraphics[width=1\linewidth]{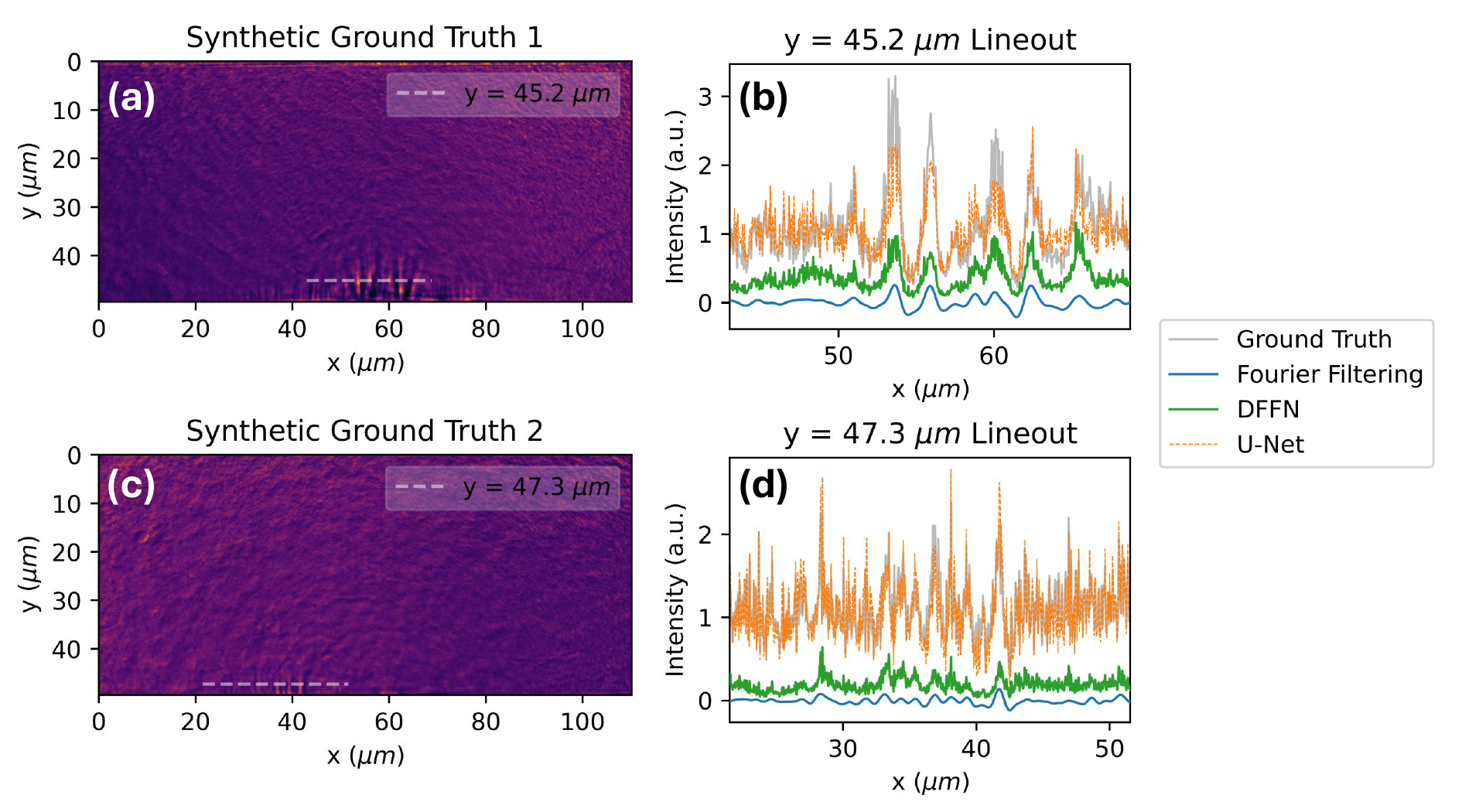}
    \caption{\textbf{(a)} Synthetic ground truth 1 (contrast-enhanced filament map with stronger filament contrast). The dashed line marks the lineout location at $y=45.2~\mu\mathrm{m}$ from the sample surface. \textbf{(b)} Lineout comparison at $y=45.2~\mu\mathrm{m}$ across the filament region, showing the ground truth profile and the profiles obtained from Fourier Filtering, DFFN, and U-Net reconstruction. \textbf{(c)} Synthetic ground truth 2 (contrast-enhanced filament map with weaker filament contrast). The dashed line marks the lineout location at $y=47.3~\mu\mathrm{m}$ from the sample surface. \textbf{(d)} Lineout comparison at $y=47.3~\mu\mathrm{m}$ across the filament region, showing the ground truth profile and the profiles obtained from Fourier Filtering, DFFN results, and U-Net reconstruction.}
    \label{fig:injection_test}
\end{figure}

\begin{table}
    \centering
    \begin{tabular}{|c|c|c|c|c|c|c|}
        \hline
        Method & MSSIM\textsubscript{1} ($\uparrow$) & PSNR\textsubscript{1} ($\uparrow$) & MSE\textsubscript{1} ($\downarrow$) & MSSIM\textsubscript{2} ($\uparrow$) & PSNR\textsubscript{2} ($\uparrow$) & MSE\textsubscript{2} ($\downarrow$)\\
        \hline
        Raw & 0.345 & 17.0 & 0.0197 & 0.068 & 14.23 & 0.0377 \\
        \hline
        Fourier Filtering & 0.459 & 26.2  & 0.0024 & 0.295 & 23.8 & 0.0042 \\
        \hline
        DFFN & 0.500 & 18.5 & 0.0141 & 0.117 & 14.6 & 0.0346\\
        \hline
        U-Net & \textbf{0.906} & \textbf{28.6} & \textbf{0.0013} & \textbf{0.945} & \textbf{29.0} & \textbf{0.0012} \\
        \hline
    \end{tabular}
    \caption{Comparisons of the mean structural similarity index (MSSIM), Peak Signal-to-Noise Ratio (PSNR) and Mean Squared Error (MSE) metrics inside the filament region of interest resulting from Fourier Filtering, DFFN and U-Net reconstruction, for the two different injection tests. The number subscript in each metric labels denotes the corresponding ground-truth image number. For MSSIM and PSNR, the higher values are better. For MSE, the lower values are better. "Raw" denotes the synthetic images.}
    \label{tab:injection test}
\end{table}

To evaluate whether the U-Net correction preserves the filament signal, we performed injection tests using two synthetic shots. The first case, as shown in Fig.~\ref{fig:injection_test}(\textbf{a}), is constructed from a contrast-enhanced filament map with relatively strong filament contrast. A second synthetic image was generated with the same method using a different filament map with weaker filamentation. In both cases, each filament map was extracted from a laser-driven shot excluded from model training and was inserted into a randomly selected cold-shot image that was likewise excluded from training. Because these tests are intended as a signal-preservation sanity check rather than a full transmission reconstruction, the injections were performed directly in the image (intensity) domain. Let $I_{cold}(x,y)$ denote the cold-shot image and $S_{fil}(x,y)$ the contrast-enhanced filament map. We constructed the synthetic image as
\begin{equation}
    I{syn} = I_{cold}(x,y) S_{fil}(x,y),
\end{equation}
We then processed each synthetic image using Fourier filtering, DFFN, and the trained U-Net model.

We quantified the filament signal preservation by computing the mean structural similarity index (MSSIM), peak signal-to-noise ratio (PSNR), mean squared error (MSE) between the injected filament ground truth and the recovered filament component within filament ROI\cite{Wang2004SSIM}. Quantitatively, the U-Net shows substantially stronger agreement with the injected filament ground truth in both synthetic cases. For the synthetic image with stronger filament contrast (synthetic ground truth 1), Fourier filtering, DFFN, and U-Net yielded MSSIM values of 0.459, 0.500, and 0.906, PSNR values of 26.2, 18.5, and 28.6, and MSE values of 0.00242, 0.0141, and 0.00138 respectively. For the synthetic image with weaker filament contrast (synthetic ground truth 2), the mean MSSIM were 0.295, 0.117, and 0.945 and the corresponding PSNR and MSE were 23.8, 14.6 and 29.0 and 0.00418, 0.0346, and 0.00125 respectively. Overall, these metrics indicate that U-Net outperforms Fourier filtering and DFFN in preserving filament signal, as evidenced by its higher MSSIM and PSNR and lower MSE in both synthetic injection tests.

In addition to MSSIM, we evaluated amplitude preservation using one-dimensional lineouts across the filament ROI. A lineout at $y=45.2\mu m$ was extracted across representative filament structures on synthetic ground truth 1 (Fig.~\ref{fig:injection_test}\textbf{(b)}) and second lineout at $y=47.3\mu m$ was extracted across representative filament structures on synthetic ground truth 2. For both lineouts, we compared the injected ground-truth profile with the recovered profiles obtained using Fourier filtering, U-Net reconstruction and DFFN. The U-Net reconstructed profile reported agreement in peak-to-background contrast and profile shape. In contrast, the Fourier filtered profile, although it retaining much of the overall profile shape, was attenuated. The DFFN profiles remain closer to the uncorrected synthetic data and retain several artifact-induced amplitude peaks.

\begin{figure}
    \centering
    \includegraphics[width=1\linewidth]{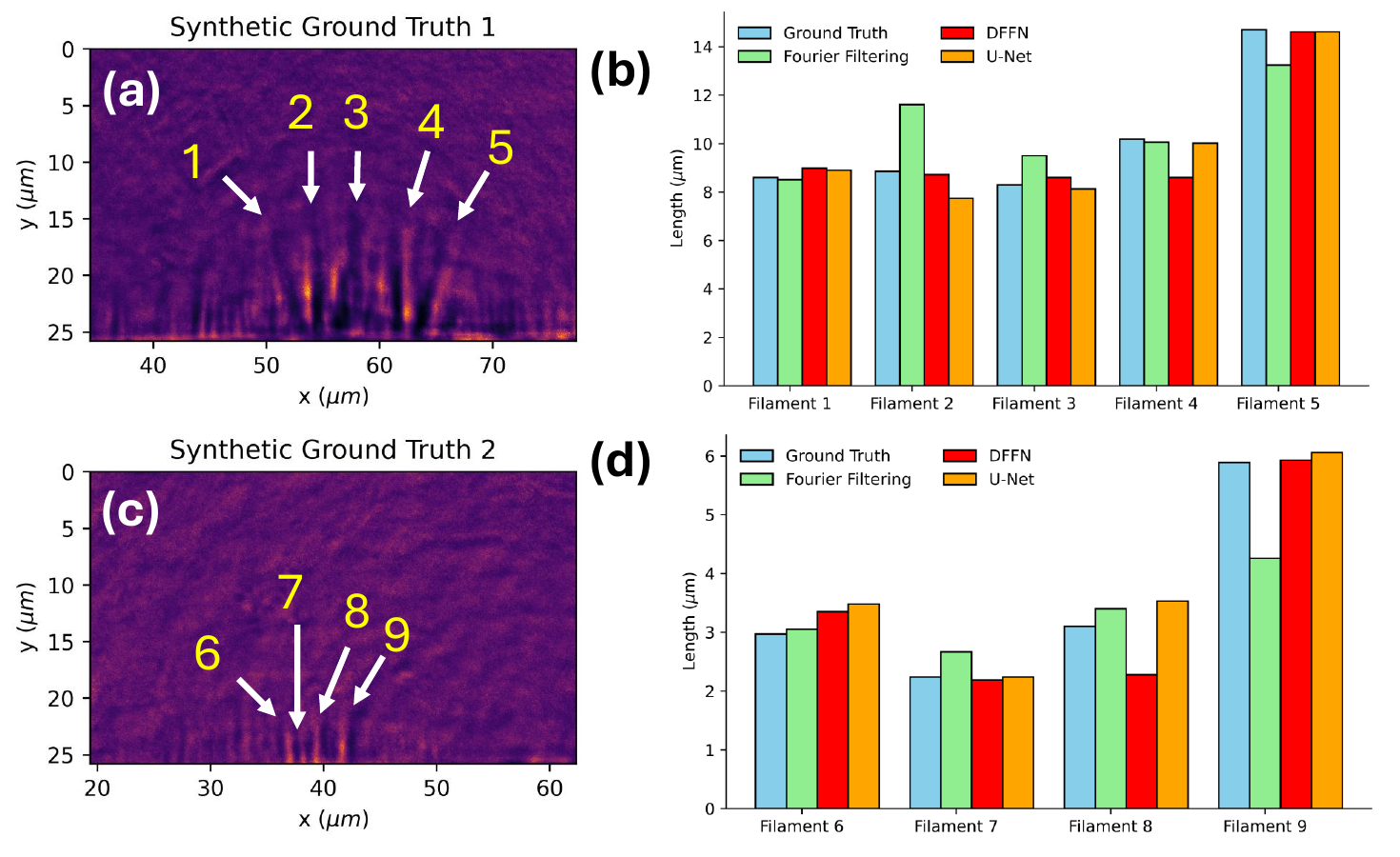}
    \caption{\textbf{(a)}, \textbf{(c)} Synthetic ground truths with the labeled filaments used for length measurement. \textbf{(b)}, \textbf{(d)} Comparison of filament lengths measured from the ground truth, U‑Net reconstruction, Fourier filtering and DFFN.}
    \label{fig:fil_length}
\end{figure}

To demonstrate the importance of filament signal preservation to quantitative measurements, we also measured the lengths of the filaments using the synthetic ground truth image (Fig.~\ref{fig:fil_length}\textbf{(a)}). We compared the recovered filament lengths after each correction method with the filaments measured in the synthetic ground truths. For each filament, we extracted the lineout along the filament direction and averaged over the filament width. The filament endpoint was then defined by the maxima or minima of the lineout, depending on the filament polarity. As shown in Fig.~\ref{fig:fil_length}\textbf{(b)}, the U‑Net reconstruction yields filament-length measurements that are consistently closer to the ground truth, whereas Fourier filtering tends to underestimate or overestimate several filament lengths because of the signal attenuation near the filament tips. Quantitatively, the root-mean-square percentage error (RMSPE) of the filament-length measurement is 8.66$\%$ for the U-Net reconstruction, compared with 16.71$\%$ for Fourier filtering and 11.3$\%$ for DFFN. This result shows U-Net reconstruction suppresses the imaging artifacts and also provides low-error filament length measurement, which is important for subsequent measurements such as electron velocity.

\subsection*{Generalizability: Shock Wave}

\begin{figure}
    \centering
    \includegraphics[width=1\linewidth]{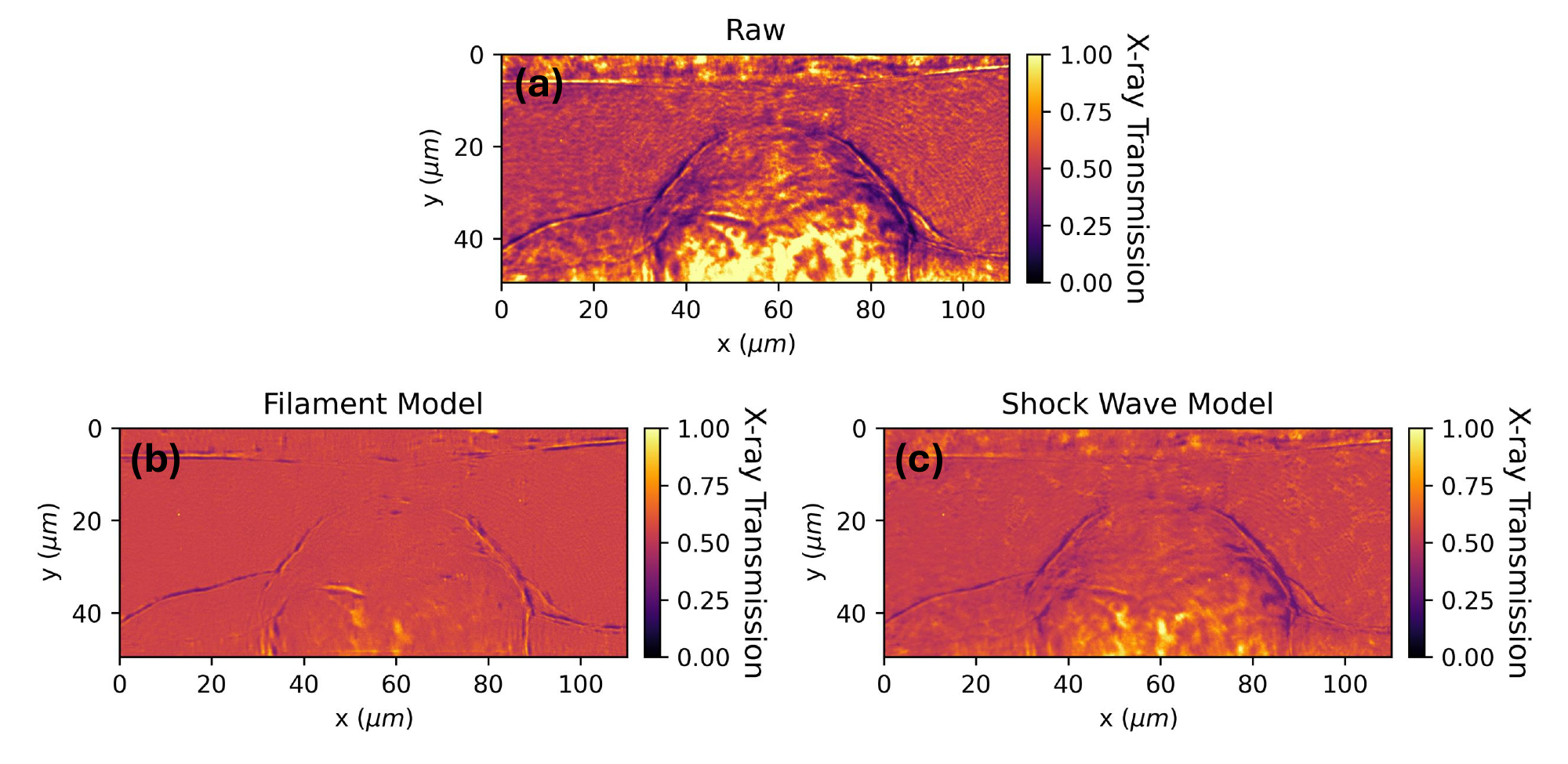}
    \caption{\textbf{(a)} Raw X-ray Transmission map of a shock wave. \textbf{(b)} Reconstructed X-ray transmission map using the filament model. \textbf{(c)} Reconstructed X-ray transmission map using the shock wave model.}
    \label{fig:shock_wave}
\end{figure}

The same model (trained with filament patches) was applied to an image taken at the later time frame, in which the early-time filamentation evolves into a shock wave. Compared to filaments, the shock wave is a large-scale object that occupies a substantial fraction of the image and can
 locally deform the feature layer (Fig.~\ref{fig:shock_wave}\textbf{(a))}. This makes the separation of physical signal and structured artifacts more challenging than in the filament regime.

 As shown in Fig.~\ref{fig:shock_wave}\textbf{(b)}, the filament-trained model does not fully preserve the shock-wave signal. Part of the shock structure is absorbed into the predicted feature layer, leading to attenuation of the reconstructed shock wave. Quantitatively, the MSSIM and PSNR between the raw and reconstructed image within the shock ROI are 0.629 and 26.1, respectively. This result indicates limited out-of-distribution generalization when the physical structure differs substantially from the filament structures represented during training.

To improve the preservation of the shock wave signal, we trained a second model with increased channels (48 initial channels). In addition, similar to the filament model, we employed the feature augmentation technique from above by dividing the shock wave into six patches and randomly pasting one of these patches onto the cold-shot images during training. As shown in Fig.~\ref{fig:shock_wave}\textbf{(c)}, this shock-aware model better preserves the shock wave structure while suppressing the feature layer, increasing the MSSIM to 0.708 and PSNR to 28.5 within the shock ROI.

\subsection*{Uncertainty Quantification via Deep Ensembles}

While deep learning offers powerful capabilities for separating signal from background, these models are inherently data-driven and susceptible to domain shift. Unlike analytical methods (such as Fourier filtering), which apply fixed mathematical operations regardless of input content and are thus universal in their applicability, a neural network optimizes its performance specifically on the statistical distribution of its training data. Consequently, ML models can become brittle when determining how to process features that were not trained to recognize, a scenario known as out-of-distribution (OOD) generalization. In HED and IFE relevant experiments, this risk is non-trivial, where a model trained to remove imaging artifacts from filament data might erroneously classify a novel physical phenomenon (e.g., a shock wave or plasma instability) as an artifact and attempt to remove it. Therefore, deploying deep learning in experimental pipelines requires not just high accuracy, but also a mechanism to quantify reliability and robustness of the underlying ML model. We address this need by utilizing a deep ensemble approach, which provides a predictive uncertainty map to flag regions where the model encounters novel physics outside its training distribution.

To assess the reliability of the artifact suppression and detect potential OOD features, we implemented a deep ensemble approach\cite{lakshminarayanan2017deepensemble}. We trained $M=10$ independent U-Net models, where each model was initialized with a different random seed and trained with different random shuffles of the data loader. This randomization between the individual models in the ensemble is to ensure de-correlation amongst the models. 

For a given input image $x$, the ensemble prediction is defined as the mean of the individual model outputs. The variability among the models captures the epistemic uncertainty (the model's lack of knowledge in regions where the data significantly differs from the training distribution). We quantify this pixel-wise uncertainty using the differential entropy of the predictive distribution. Assuming the ensemble predictions follow a Gaussian distribution, the entropy $h(X)$ at each pixel is given by\cite{Sepulveda2024entropy}:
\begin{equation}
    h(X) = \frac{1}{2} \log(2\pi e \sigma^2),
\end{equation}
where $\sigma^2$ is the ensemble variance of the X-ray transmission map.

\begin{figure}
    \centering
    \includegraphics[width=1\linewidth]{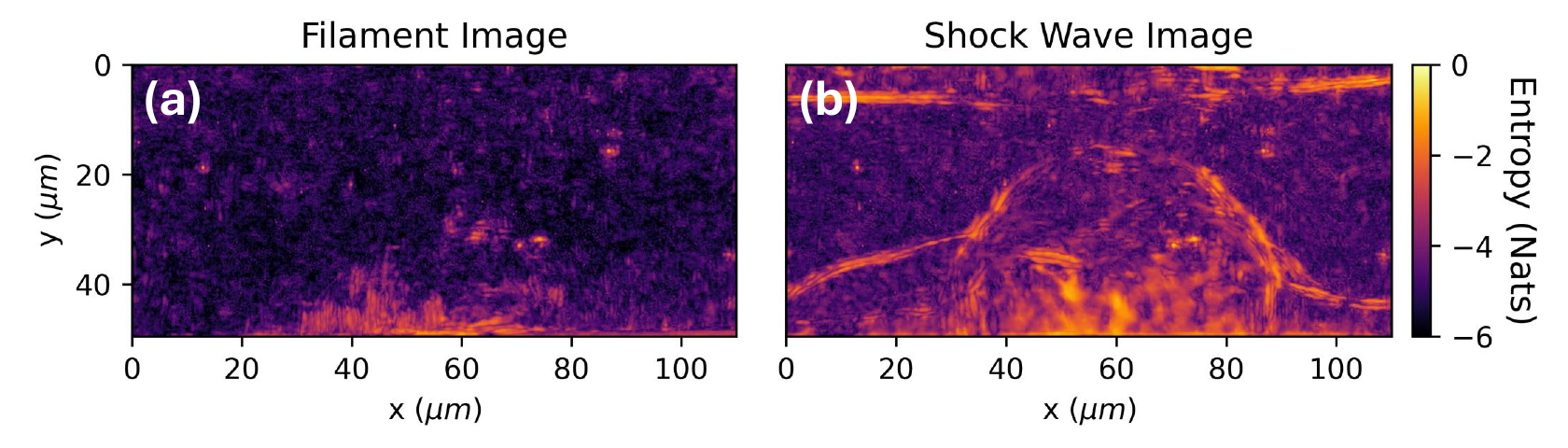}
    \caption{Pixel-wise entropy map computed from an ensemble of 10 models for \textbf{(a)} a filament image and \textbf{(b)} a shock wave. Brighter color (higher entropy) represents higher uncertainty, highlighting regions where the ensemble shows the largest disagreement (typically near the prominent physical structures).} 
    \label{fig:Ensemble}
\end{figure}

This high entropy serves as an effective OOD detector. It signals that the shock wave is physically distinct from the imaging artifacts the network was trained to remove. Consequently, the high variance regions in the reconstructed transmission map should be interpreted with caution, as the model lacks the learned features to cleanly separate the signal from the background in this regime. It should be noted that the entropy maps are shown on a logarithmic scale. In the filament map, the filaments appear less bright because of the filament augmentation approach. However, they remain brighter than the background, which is still sufficient for flagging physical structures.

\section*{Discussion}

The artifacts from the imaging system are structured and drift shot to shot rather than spatially uncorrelated noise. They prevent clean cancellation in flat-field normalization and imprint the residual into the X-ray transmission map, limiting the accuracy of quantitative filament measurements.

Fourier filtering is limited in our data because the artifacts are not confined to isolated frequencies, and the artifacts overlap with the filament signal in the Fourier domain. Aggressive filtering suppresses the artifacts but can also attenuate filament signal and distort filament profiles. Unlike the Fourier-filtering method, U-Net treats the artifacts as a separable feature layer on a per-image basis and therefore does not require perfect alignment between the laser-driven image and flat-field image. 

As an additional baseline, we evaluated DFFN. In our dataset, this approach produces only small modifications to the flat field, and the structured artifacts remain in the transmission map. This result is mainly due to chromatic aberration, in which the magnification and focus of the artifacts vary from shot to shot because of shifts in the XFEL central photon energy and spectral profile.

We verified the signal preservation of U-Net reconstruction with two synthetic injection tests and found strong agreement between the injected filament ground truths and the recovered filaments within the ROI. Lineouts across the filament ROI show that the U‑Net reconstruction maintains filament amplitude and profile shape after artifact suppression. In contrast, Fourier filtering yields lower MSSIM and PSNR and higher MSE in both injection tests, suggesting that the artifact suppression is achieved partly by attenuating filament signal rather than by removing only the artifact. This is consistent with the limitation of Fourier filtering in our dataset: suppressing the structured artifact requires masking Fourier components that also contribute to the filaments, which can reduce filament visibility and distort filament profiles. The filament-length measurements further confirm that filament signal preservation is essential for quantitative interpretation. Fourier filtering attenuates the tips of the filament, which leads to larger errors than U-Net reconstruction. 

When the model is applied to a structure from a different domain, the model may misclassify unfamiliar structures as artifacts. For example, the shock wave is a large-scale, complex structure that occupies a substantial fraction of the image and can locally alter the artifact pattern. As a result, the filament-trained model can partially absorb the shock wave into the predicted feature layer, leading to attenuation of the reconstructed shock structure. This suggests that the model separates artifacts and filament-like structures effectively, but does not necessarily generalize to substantially different structures. In this regime, training a shock-aware model with increased capacity and shock-patch augmentation improves preservation of the shock signal (MSSIM in the shock ROI increases from 0.629 to 0.708). This demonstrates that the feature-layer removal framework remains adaptable, but the U-Net model benefits from including examples of new structures during training.

Because learning-based artifact suppression is inherently data-driven, robust deployment of the model in experimental pipelines requires not only high accuracy on in-distribution data but also a mechanism to detect when the model encounters unfamiliar structures. We addressed this using a deep-ensemble approach ($M=10$ independently trained U-Net models), where disagreement among ensemble models provides an estimate of epistemic uncertainty. The resulting pixel-wise entropy map highlights regions where model predictions disagree, providing a "confidence layer" that can flag novel physics or unreliable reconstruction.

It should be noted that substantial changes to the beamline configuration and imaging optics can modify the artifacts and reduce model performance. In these cases, the network should be retrained or fine-tuned with reference data from the new setup. 

In conclusion, we present a U-net based method to suppress the structured, non-stationary artifacts introduced by the MEC X-ray imaging system that enables robust reconstruction of X-ray transmission maps. By applying the same model to both the laser-driven and flat-field images prior to normalization, the method reduces residual artifact imprint while preserving filament signal, as demonstrated by the injection tests and filament length measurements. These capabilities enable more reliable quantitative extraction of transport observables in HED experiments, a key capability for the IFE effort. Moreover, because the model is trained once offline and then applied to new images through a single forward pass, the approach is well suited to the large datasets expected from high-repetition-rate X-ray beamlines.

\section*{Methods}

\subsection*{Models}

\subsubsection*{Data processing}

\begin{figure}
    \centering
    \includegraphics[width=\linewidth]{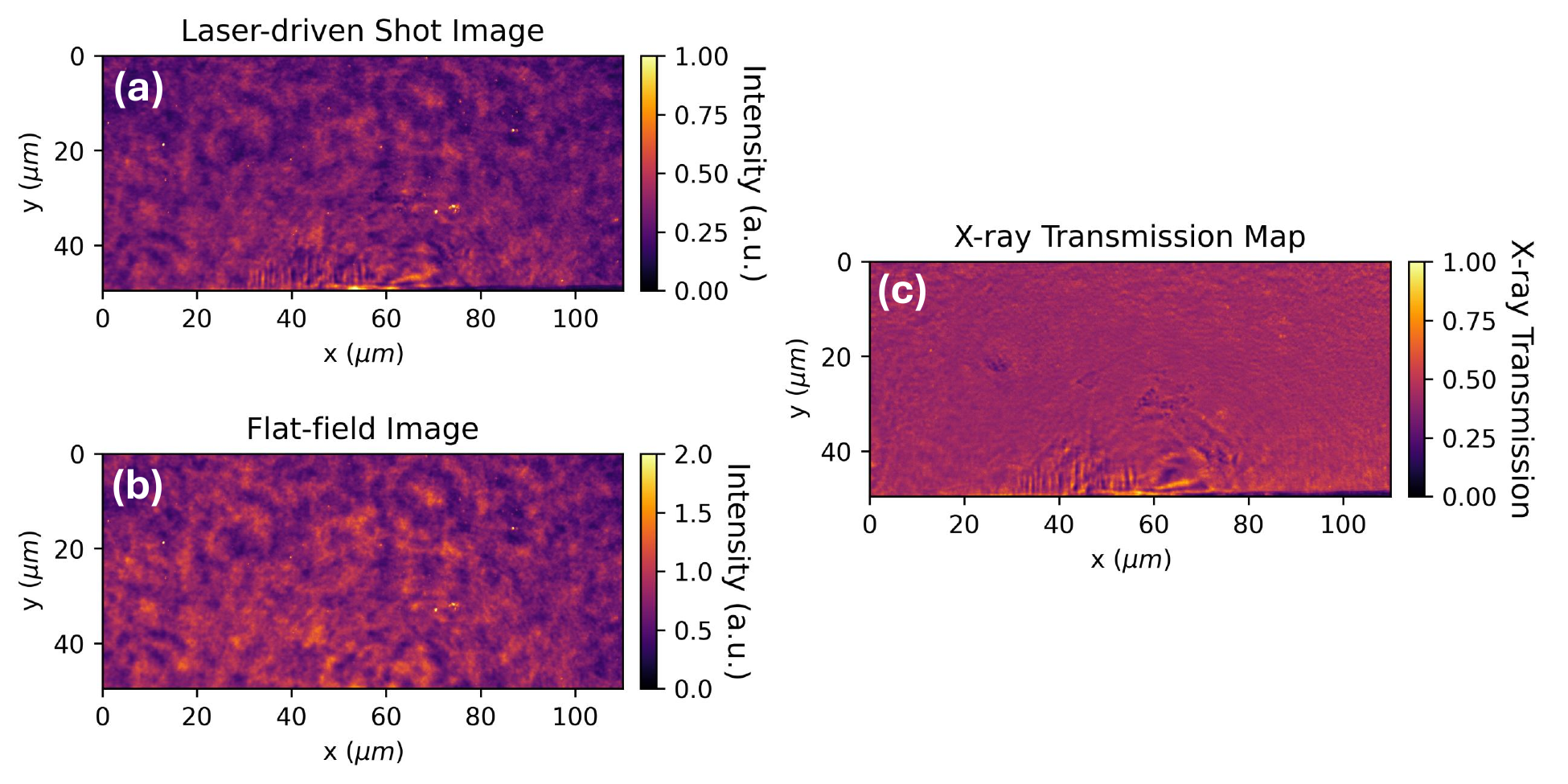}
    \caption{\textbf{(a)} Laser-driven shot image with filaments formed on the bottom of the image. Laser hit the Si bulk sample from the bottom. \textbf{(b)} Flat-field image. \textbf{(c)} X-ray transmission map computed using the images from \textbf{(a)} and \textbf{(b)}.}
    \label{fig:data_processing}
\end{figure}
The images were recorded using the MEC X-ray imager (MXI)\cite{Galtier2025MXI}. For each laser shot, we reconstructed an X-ray transmission map by normalizing the laser-driven image $I_{shot}$ by a flat-field reference $I_{flat}$. Prior to flat-field normalization, each image was 90th-percentile-normalized (to avoid scaling based on hot pixels) to place the images on the same intensity scale. After flat-field normalization, the mean X-ray transmission in the unperturbed in-sample region (away from the laser-driven interaction) was approximately 42~$\%$, which served as a consistency check for the transmission reconstruction\cite{Henke1993CXRO}. 
Fig.~\ref{fig:data_processing} shows an example of a raw laser-driven image, the corresponding flat-field reference, and the resulting X-ray transmission map. As can be seen, the structured features caused by the defects in imaging system are visible in both $I_{shot}$ and $I_{flat}$. Because these features drift slightly from shot to shot (due to X-ray beam pointing and photon-energy variations), their imperfect cancellation leaves residual patterns in $T(x,y)$ that can spatially overlap with the filament signal.

\subsubsection*{Fourier Filtering}

As a baseline suppression method, we applied Fourier-domain filtering to $I_{shot}$ and $I_{flat}$ separately prior to transmission reconstruction\cite{Munch2009FFiltering}. For each image, we computed the two-dimensional fast Fourier transform (FFT) and suppressed low-frequency background variations using a central low-frequency mask of 20 pixels. We then applied an additional 99.5th-percentile magnitude thresholding to suppress the structured feature layer. This cutoff was selected empirically to maximize feature suppression while minimizing removal of the signal of interest. The corrected transmission map was obtained by computing the inverse FFT. Because Fourier filtering can modify the overall intensity offset, we restored the global mean by adding a constant offset equal to the difference between the mean of the original image and that of the filtered image. The corrected X-ray transmission map was then reconstructed by dividing the corrected laser-driven image by the corrected flat-field image.

\subsubsection*{Dynamic Flat-Field Normalization}

DFFN was used to correct the dynamics in flat field before normalization\cite{VanNieuwenhove2015flatfield}. Let $I_{flat,m}$ denote the $m$th flat-field image and $\bar{I}_{flat}$ the mean flat field. Each flat field was approximated as a linear combination of $K$ eigen flat fields (EFFs),
\begin{equation}
    I_{flat,m} \approx \bar{I}_{flat} + \sum_{k=1}^{K} w_{mk} u_k,
\end{equation}
where $u_k$ are the EFFs and $w_{mk}$ are the corresponding weights.
The EFFs can be computed using Principal Component Analysis (PCA). The number of retained EFFs was determined by parallel analysis. In this work, $S=100$ synthetic flat-field matrices with the same dimensions and per-pixel variance as the laser-driven shot image were generated, and component $k$ was retained only when its eigenvalue exceeded the 95th percentile of the corresponding eigenvalues of the sampled matrices. The weights of the chosen flat fields is calculated by minimizing the total variation in the laser-driven image. 
For each laser-driven image, its flat field was estimated as 
\begin{equation}
    \hat{I}_{flat} = \bar{I}_{flat} + \sum_{k=1}^{K} \hat{w}_{k} u_k.
\end{equation}
The coefficients $\hat{w}_{jk}$ were obtained by minimizing the total variation of the X-ray transmission map,
\begin{equation}
    T_j(x,y;\mathbf{w}_j) = \frac{I_{shot,j}(x,y)}{\bar{I}_{flat}(x,y)+\sum_{k=1}^{K} w_{jk}u_k(x,y)},
\end{equation}
To avoid a trivial low-intensity solution, the total variation term was multiplied by the mean of the estimated flat field. The objective function was minimized using a quasi-Newton method with a tolerance level of $10^{-6}$ and 400 iterations. Because the artifacts overlap spatially with the filaments, the filament ROI was masked to prevent the filaments from affecting the minimization process. The resulting flat field was then used to normalize the laser-driven image. 

\subsubsection*{U-Net}

We used a two-dimensional U-Net (see Methods: U-Net). Two cold-shot datasets (silicon sample present, X-ray probe only; no optical drive) were used for model training, and one additional cold-shot dataset were used for validation. To increase the training speed, all images were pre-normalized to range of $[0,1]$. 
To discourage the network from interpreting filaments as artifacts, we employed an augmentation approach in which filament patches cropped from two laser-driven shots were randomly overlaid onto training images with 90 $\%$ probability at random locations\cite{ghiasi2021dataaugmentation}.

The feature layer of the raw data is suppressed by dividing the raw data by the prediction of the U-Net, i.e., $I_{clean}=I_{raw}/I_{prediction}$. To preserve the global mean such that the corrected X-ray transmission is still close to the real physical value, we multiplied the cleaned data by the mean of the raw data. We applied the same trained model to the flat-field. The X-ray transmission is then reconstructed by dividing the corrected laser-driven image and the flat-field.

\subsection*{Experiment Parameters}

The experiment was performed at the Matter in Extreme Conditions (MEC) instrument of the Linac Coherent Light Source (LCLS) on 100-$\mu m$-thick (in the X-ray direction) silicon targets, irradiated by a 1 J, 45 fs, 7 $\mu m$ (Full Width at Half Maximum) focus, 800 nm optical laser and probed by 9.5-keV XFEL pulses. The XFEL pulses were delayed relative to the optical laser to capture interactions at different time frames. 

\subsection*{U-Net} \label{section: Methods/U-Net}

\begin{figure}
    \centering
    \includegraphics[trim={1.55cm 0 0.25cm 0},clip, width=1.025\textwidth]{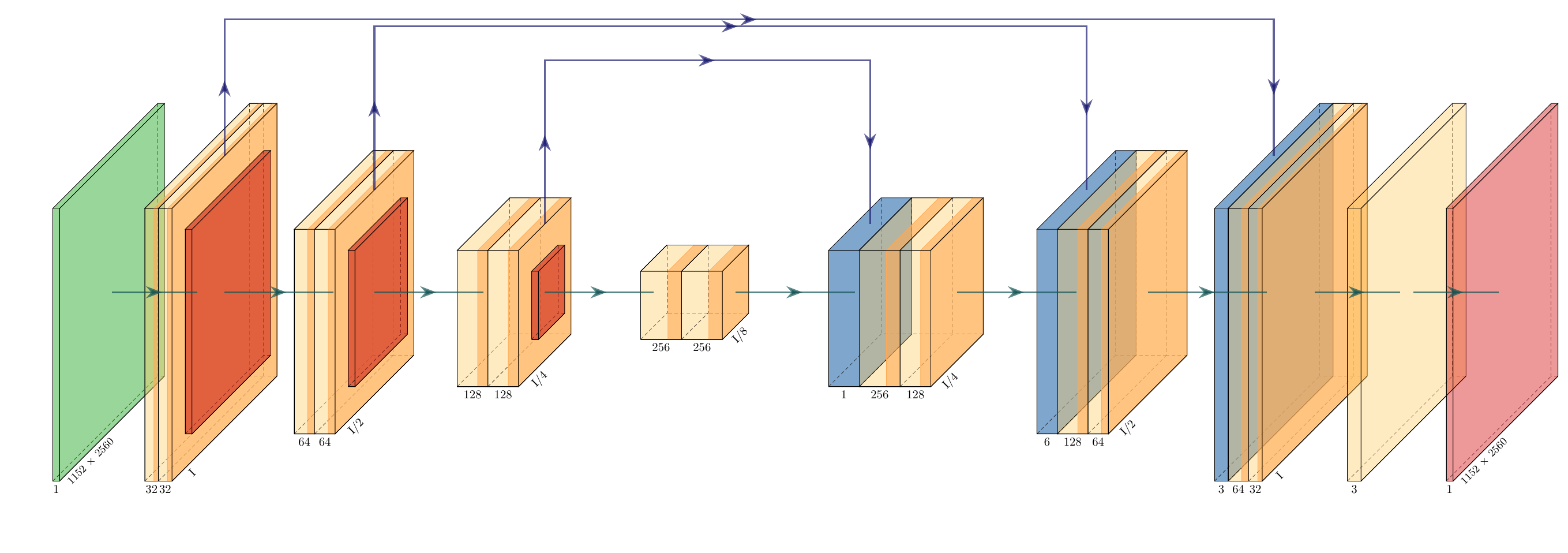}
    \caption{U-Net architecture diagram of the model used in this study. The Convolution layers are shown in yellow, the pooling layers in deep red and the upsampling layers in blue. The input is denoted in green and the final output in red. The Encoder consists of three blocks, each blocks is composed from convolution-convolution-pooling operations. The Decoder mirrors this by using three blocks, composed of upsampling-convolution-convolution operations. The dimensions of each feature map are labeled, where $I$ refers to the input feature map's height and width.}
    \label{fig:U-Net schematic}
\end{figure}

The U-Net presented in this paper consists of three downsampling stages and a symmetric decoder, corresponding to four spatial scales (full resolution, /2, /4, and /8). The encoder consisted of three convolutional blocks with 32, 64, and 128 feature channels, followed by a bottleneck block with 256 channels. Each block contained two $3 \times 3$ convolutions with padding of 1 and GELU activations~\cite{hendrycks2016GELU}. Downsampling was performed using $2 \times 2$ max pooling, and upsampling used $2 \times 2$ transposed convolutions (stride 2). Skip connections were implemented by concatenation between encoder and decoder feature maps at matching resolutions. A final $1 \times 1$ convolution produced a single-channel output. The network contained 1925025 parameters. 

To prevent the network from absorbing filaments into the predicted feature layer, we used a filament-aware weighted L1 loss during training. The loss function $\mathcal{L}$ is defined 
\begin{equation}
    \mathcal{L} = \frac{1}{N} \sum_{i=1}^{N} (1 + \alpha M_i) | \hat{I}_i - I_i |
\end{equation}
with $\alpha=10$, corresponding to an $11\times$ weight on pixels within pasted regions and $M(x,y) \in \{0,1\}$ represents the binary mask of copy-paste augmentation (1 inside pasted filament patches and 0 elsewhere). Two cold-shot datasets consisting of a total 197 images were used for training. Each image originally had a size of 2160 $\times$ 2560 pixels and was normalized by its 90th-percentile value to place all images on the same intensity scale. The images were then cropped to 1152 $\times$ 2560 pixels so that the training data contained only the sample ROI. A third cold-shot dataset consisting of 100 images was used for validation. We trained the model using an Adam optimizer with a learning rate of $10^{-4}$ and default $\beta$ parameters ($\beta_1=0.9$, $\beta_2=0.999$)\cite{kingma2014adam}. The model was trained for 20 epochs.

As previously mentioned, the artifacts behave as a multiplicative modulation. We therefore converted the training and validation images to the logarithmic domain to facilitate model training. In the logarithmic domain, the model was trained to predict the artifact layer. The artifact layer was then converted back to intensity space. The artifacts were suppressed by dividing the raw data image with the predicted artifact layer. The laser-driven image and flat-field image were reconstructed separately using the same trained model. Before reconstructing the X-ray transmission map, each corrected laser-driven and flat-field image was multiplied by the mean pixel value of its corresponding raw image to preserve the mean value of the reconstructed X-ray transmission map.

\subsection*{Enhanced contrast feature extraction}

Two laser-driven images were selected for filament patch extraction. For each selected shot, we computed the mean of the corresponding cold-shot sets, $I_{cold,mean}$ and converted both the $I_{shot}$ and $I_{cold,mean}$ to log domain, $X_{shot}=log(I_{shot}+\epsilon)$ and $X_{cold,mean}=log(I_{cold,mean}+\epsilon)$, where $\epsilon$ is a small constant to avoid $log(0)$. To compensate for shot-to-shot drift of the structured artifacts, we estimated the relative shift between $X_{cold,mean}$ and $X_{shot}$ using the artifact pattern within the filament ROI. To avoid biasing the alignment using filament structures, the translation was computed by phase correlation using only the upper portion of the ROI, which does not contain filaments. We then shifted $X_{cold,mean}$ into $X_{shot}$ coordinates and formed an artifact-suppressed filament residual by subtraction, $R =X_{shot} -X_{cold,mean}$. Filament patches were cropped from R and normalized to the range [-1,1].

\section*{Acknowledgements}

This work is supported by the Department of Energy Laboratory Directed Research and Development (LDRD) program at SLAC National Accelerator Laboratory under contract number DE-AC02-76SF00515. SLAC National Accelerator Laboratory is supported by the U.S. Department of Energy, Office of Science, under Contract No. DE-AC02-76SF00515. Use of the Linac Coherent Light Source (LCLS) is supported by Basic Energy Sciences; use of the MEC Instrument is further supported by Office of Fusion Energy Sciences under FWP 100106.

\section*{Funding}

This work is supported by the Department of Energy Laboratory Directed Research and Development (LDRD) program at SLAC National Accelerator Laboratory under contract number DE-AC02-76SF00515. SLAC National Accelerator Laboratory is supported by the U.S. Department of Energy, Office of Science, under Contract No. DE-AC02-76SF00515. Use of the Linac Coherent Light Source (LCLS) is supported by Basic Energy Sciences; use of the MEC Instrument is further supported by Office of Fusion Energy Sciences under FWP 100106.

\section*{Author contributions statement}

S.X.L. and A.A.M. performed the coding and the data driven modeling. E.G. designed the laser experiment. E.G., A.A., M.B., N.B., G.C., E.C., N.C., G.D., J.E., B.E., A.G., M.G., P.H., D.K., H.J.L., P.M., B.N., P.N., C.R., M.W. and A.Z. conducted the laser experiment. S.X.L. and A.A.M. wrote the paper. E.G. and Q.L.N. provided supervision. All authors reviewed the manuscript. 

\section*{Data availability statement}
The data used in this study will be made available upon reasonable request. 

\section*{Code availability statement}
The code used in this study will be made available as a Github repository upon publication.
\bibliography{sample}

\end{document}